\documentclass[12pt, a4paper]{article}

\usepackage{amsmath,amssymb,bm,epsfig,afterpage}
\usepackage{cite}
\usepackage{here}
\usepackage{color}
\usepackage{colortbl}
\usepackage{xcolor}
\usepackage{tabularx}
\usepackage{subcaption}
\usepackage{bbm}
\usepackage{graphicx}
\usepackage{listings}
\usepackage{longtable}
\usepackage[utf8]{inputenc}
\usepackage{cancel}
\usepackage{slashed}

\allowdisplaybreaks[3]

\newcommand{\ka}{\kappa}

\newcommand{\kap}{{\kappa^\prime}}
\newcommand{\Zp}{{Z^\prime}}
\newcommand{\gp}{{g^\prime}}
\newcommand{\Qp}{Q^\prime}
\newcommand{\ve}{\mathbf{e}}

\newcommand{\GeV}{\mathrm{GeV}}
\newcommand{\TeV}{\mathrm{TeV}}

\newcommand{\la}{\lambda}
\newcommand{\eps}{\epsilon}

\newcommand{\Mcal}{\mathcal{M}}
\newcommand{\Lcal}{\mathcal{L}}

\newcommand{\ev}{\mathbf{e}}

\newcommand{\SM}{{\mathrm{SM}}}

\newcommand{\ol}[1]{\overline{#1}}

\newcommand{\vev}[1]{\langle{#1}\rangle}

\newcommand{\br}[2]{\mathrm{BR} \left({#1}\to{#2}\right)}

{

\newcommand{\expr}{{\mathrm{exp}}}

\newcommand{\bsll}{b\to s\ell\ell}
\newcommand{\damu}{\Delta a_\mu}
\newcommand{\Up}{U(1)^\prime}

\newcommand{\SRtt}{\texttt{SR0}^{\mathrm{tight}}_{\mathrm{bveto}}}

\newcommand{\val}[2]{{#1}\times 10^{#2}}

\usepackage[colorlinks=true, linkcolor=black, citecolor=black,
urlcolor=black]{hyperref}

\setcounter{footnote}{0}

\newif\iffigsame
\figsametrue

\begin{document}
\begin{flushright}
 {\tt
CTPU-PTC-22-02  
}
\end{flushright}
\vspace{0.0cm}
\begin{center}
{\Large
{\bf
Complete vectorlike fourth family and new U(1)$^\prime$ \\ for muon anomalies~\footnote{
Invited talk at 20th Lomonosov Conference, Moscow, Russia, August 21, 2021. 
}
}
\vskip 0.5cm
}
Junichiro Kawamura~\footnote{jkawa@ibs.re.kr}
\vskip 0.1cm
{\it 
Center for Theoretical Physics of the Universe, Institute for Basic Science (IBS),
Daejeon 34051, Korea
}\\[3pt]
\vskip 0.3cm
\begin{abstract}
I introduce a model to explain the recent anomalies in the muon anomalous magnetic moment 
and the rare semi-leptonic $B$ meson decays.  
The Standard Model is extended by a $U(1)^\prime$ gauge symmetry 
and a complete fourth family fermions which are vector-like under the gauge symmetry. 
We found parameter points which can explain the anomalies 
consistently with the other observables. 
We then propose an interesting possibility to have signals with four muons or more 
from vector-like lepton pair production. 
\end{abstract}
\end{center}

\section{Introduction}

The discrepancies from the Standard Model (SM) predictions 
are found in the measurements of 
the muon anomalous magnetic moment   
and the rare semi-leptonic decays of $B$ mesons, $\bsll$. 
The latest measurements of $\Delta a_\mu = a_\mu^\expr-a_\mu^\SM$ 
and the lepton universality observable, $R_K$, confirmed 
that their values are deviated from the SM prediction 
by $4.2\sigma$~\cite{Aoyama:2020ynm,Abi:2021gix} 
and $3.1\sigma$~\cite{LHCb:2021trn}, respectively. 
These anomalies may imply that 
there are new particles at the TeV scale which couple to muons.

In our papers~\cite{Kawamura:2019rth,Kawamura:2019hxp}, 
we proposed a model which explains the anomalies 
in $\damu$ and $\bsll$ simultaneously~\footnote{
Similar models for the anomalies are studied in Refs.~\cite{Allanach:2015gkd,Altmannshofer:2016oaq,Megias:2017dzd,Raby:2017igl,Arnan:2019uhr}
}.   
We introduced a complete vector-like (VL) family of quarks and leptons, 
and a new vector-like $\Up$ gauge symmetry.  
Only the VL family and the $\Up$ breaking scalar, $\Phi$,  
carry non-zero $\Up$ charges,  
whereas the three generations of chiral families and the Higgs boson are neutral. 
A $\Zp$ boson associated with the $\Up$ gauge symmetry 
couples to the SM families via the mixing between the chiral and VL families 
induced by the $\Up$ breaking.  
In this model, $\damu$ is explained by the 1-loop contributions mediated 
via the VL leptons and the $\Zp$ boson, 
and $\bsll$ is explained by the tree-level exchanging of the $\Zp$ boson. 
We showed that the anomalies can be explained 
if the VL leptons and $\Zp$ boson are lighter than 1.5 TeV.

We propose a novel possibility 
to search for the VL leptons and $\Zp$ boson~\cite{Kawamura:2021ygg}. 
We consider the pair production of VL leptons decaying to a $\Zp$ boson 
followed by the $\Zp$ boson decaying to a pair of muons or muon neutrinos. 
There are six (four) muons in the final state 
if both (either of) the $\Zp$ bosons decay to muons.  
We recast the latest ATLAS search~\cite{Aad:2021qct} 
for events with four or more charged leptons.

\section{Explanation for the muon anomalies}
\label{sec-model}

\subsection{Model} 
\begin{table}
\centering
\vspace{0.5cm}
\caption{\label{tab-mcEXO}
Quantum numbers of new fermion and scalar fields.
}
 \begin{tabular}{c|cccccc|cccccc|cc}\hline
         & $Q_L$&$\ol{U}_R$&$\ol{D}_R$&$L_L$&$\ol{E}_R$&$\ol{N}_R$&
         $\ol{Q}_R$&$U_L$&$D_L$&$\ol{L}_R$&$E_L$&$N_L$&
         $\phi$ &$\Phi$\\ \hline\hline
 $SU(3)_{\mathrm{C}}$&$\bf{3}$&$\ol{\bf{3}}$& $\ol{\bf{3}}$&$\bf{1}$&$\bf{1}$& $\bf{1}$&
                      $\ol{\bf{3}}$&$\bf{3}$& $\bf{3}$&$\bf{1}$&$\bf{1}$& $\bf{1}$&$\bf{1}$&$\bf{1}$\\
 $SU(2)_{\mathrm{L}}$&$\bf{2}$&$\bf{1}$& $\bf{1}$& $\bf{2}$& $\bf{1}$& $\bf{1}$&
                      $\bf{2}$&$\bf{1}$& $\bf{1}$& $\bf{2}$& $\bf{1}$& $\bf{1}$&$\bf{1}$&$\bf{1}$   \\
 $U(1)_{\mathrm{Y}}$ &$1/3$&$\text{-}4/3$&$2/3$&$\text{-}1$&$2$&$0$&
                    $\text{-}1/3$&$4/3$&$\text{-}2/3$&$1$&$\text{-}2$&$0$&
                    $0$&$0$ \\ \hline
 $U(1)^\prime$     & $\text{-}1$ & $1$ & $1$ & $\text{-}1$ & $1$ & $1$ & $1$ & $\text{-}1$ & $\text{-}1$ & $1$ & $\text{-}1$ & $\text{-}1$ & $0$ & $\text{-}1$ \\ \hline
 \end{tabular}
\end{table} 

We review the model proposed in Refs.~\cite{Kawamura:2019rth,Kawamura:2019hxp}.
The new particles of our model are given in Table~\ref{tab-mcEXO}. 
The SM particles are neutral under the $\Up$ gauge symmetry. 
For simplicity, we focus on the charged leptons.  
The first and third generations of the chiral leptons are omitted,    
although we studied the model with complete generations 
and discussed lepton flavor violation.  
The mass of the VL states and Yukawa interactions are given by
\begin{align}
 \Lcal \supset&\ - m_L \ol{L}_R L_L -  m_E \ol{E}_R E_L  \\ \notag
         &\     + y_\mu \ol{\mu}_R \ell_L H
               + \kap \ol{E}_R L_L H  - \ka \ol{L}_R \tilde{H} E_L
                 + \la_L \Phi   \ol{L}_R \ell_L - \la_E \Phi \ol{\mu}_R E_L + h.c. ,
\end{align}
where $\tilde{H} := i\sigma_2 H^* = (H_-^*, -H_0^* )$.
The $SU(2)_L$ doublets are defined as
\begin{align}
\ell_L = (\nu_L, \mu_L), \quad
 H =(H_0, H_-),\quad
 L_L = (N_L^\prime, E_L^\prime), \quad
 \ol{L}_R = (-\ol{E}_R^\prime, \ol{N}^\prime_R), 
\end{align}
and the $SU(2)_L$ indices are contracted via $i\sigma_2$.
After symmetry breaking by $v_H := \vev{H_0}$ and $v_\Phi := \vev{\Phi}$,
the mass matrix for the leptons is given by
\begin{align}
 \ol{\ve}_R \Mcal_e \ve_L  :=&\
\begin{pmatrix}
 \ol{\mu}_R & \ol{E}_R & \ol{E}_R^\prime
\end{pmatrix}
\begin{pmatrix}
 y_\mu v_H & 0        & \la_E v_\Phi \\
 0         & \kap v_H & m_E \\
\la_L v_\Phi & m_L & \ka v_H
\end{pmatrix}
\begin{pmatrix}
 \mu_L \\ E^\prime_L \\ E_L
\end{pmatrix}.    
\end{align}
The mass basis is defined as
\begin{align}
 \hat{\ve}_L := U_L^\dag \ve_L,
\quad
 \hat{\ve}_R := U_R^\dag \ve_R,
\quad
 U_R^\dag \Mcal_e U_L = \mathrm{diag}\left(m_\mu, m_{E_2}, m_{E_1}\right),
\end{align}
where $E_1$ and $E_2$ are respectively the singlet-like and doublet-like VL leptons.
We define the Dirac fermions as
$\ev := \left(\mu, E_2, E_1\right)$,  
$\left[\ev\right]_i
:=\left( \left[\hat{\ev}_L\right]_i,\left[\hat{\ev}_R \right]_i \right)$,                
where $i=1,2,3$. 
The mass matrices for the quarks and the Dirac mass matrices for the neutrinos 
have the same structure as the charged leptons. 
We assume that the three generations of right-handed neutrinos have Majorana masses 
at the intermediate scale, 
so that the tiny neutrino masses are explained by the type-I see-saw mechanism.

The gauge interactions with the $\Zp$ boson in the mass basis are given by 
\begin{align}
 \Lcal_{V} =&\ Z^\prime_\mu \; 
                  \ol{\ve} \gamma^\mu \left({g}^\Zp_{e_L} P_L + {g}^\Zp_{e_R}P_R  \right)
                  \ve,
\quad 
 g^\Zp_{\ev_L} = \gp U_L^\dag \Qp_e
  U_L, \quad
 g^\Zp_{\ev_R} = \gp U_R^\dag \Qp_e
  U_R
\end{align}
where $Q_e^\prime = \mathrm{diag}(0,1,1)$ is the coupling matrix in the gauge basis. 
Note that the chiral family does not couple to the $\Zp$ boson 
in the gauge basis, and the coupling arises only in mass basis. 
Here, $P_{L}$ ($P_R$) are the chiral projections onto the left- (right-)handed fermions.
$\gp$ is the gauge coupling constant for $U(1)^\prime$.

\subsection{Muon anomalies}

\begin{figure}[t]
\begin{minipage}[c]{0.48\hsize}
 \centering
\includegraphics[height=30mm]{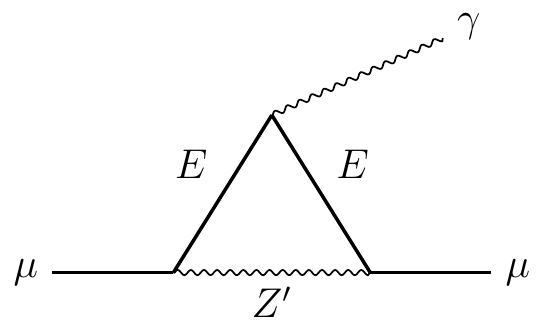}
\end{minipage}
\begin{minipage}[c]{0.48\hsize}
 \centering
    \includegraphics[height=30mm]{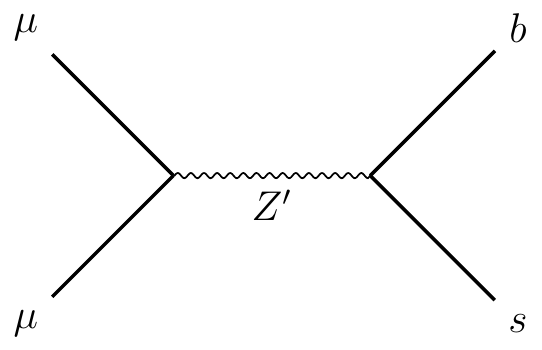}
\end{minipage}
 \caption{\label{fig-anom}
Diagrams contribute to $\Delta a_\mu$ (left) and the $b\to s\ell\ell$ decay (right). 
}
\end{figure}

The diagrams that can explain the anomalies are shown in Fig.~\ref{fig-anom}. 
The $\Zp$ boson 
and the physical mode $\chi$ of the $\Phi$ boson contribute to $\Delta a_\mu$ 
as~\cite{Jegerlehner:2009ry,Dermisek:2013gta}
\begin{align}
\label{eq-delamu}
 \Delta a_\mu \sim
    \frac{  m_\mu \ka v_H}{64\pi^2 v_\Phi^2} s_{2L} s_{2R}  
\left(\sqrt{x_L x_E} \frac{G_Z(x_L)-G_Z(x_E)}{x_L - x_E}
+ \frac{1}{2} \sqrt{y_L y_R} \frac{{y_L} G_{{S}}(y_L)-{y_R} G_{{S}}(y_R)}{y_L-y_R}
\right),
\end{align}
where $s_A := \la_X v_\Phi/M_X$,   
with $M_X^2:=m_X^2 + \la_X^2 v_\Phi^2$~($X = L,E$ for $A=L,R$), 
is the mixing angle between the chiral and VL families.  
Here, $x_L := M_L^2/m_\Zp^2$, $x_E:= M_E^2/m_\Zp^2$,
$y_L:= M_L^2/m_\chi^2$, $y_E := M_E^2/m_\chi^2$  
with $m_\Zp^2 = 2 \gp^2 v_\Phi^2$ and $m_\chi$ is the mass of $\chi$. 
The loop functions are given by
\begin{align}
  G_Z(x) :=&\  \frac{x^3+3x-6x \ln{(x)}-4}{2(1-x)^3}, \quad
  G_S(y) := \frac{y^2-4y+2\ln{(y)}+3}{(1-y)^3}.
\end{align}
The value of $\damu$ is estimated as 
\begin{align}
 \Delta a_\mu \sim 2.9\times 10^{-9} \times
                 \left( \frac{1.0~\TeV}{v_\Phi}\right)^2
                 \left( \frac{\ka}{1.0} \right)
                 \left( \frac{s_{2L}s_{2R}}{1.0} \right)
                 \left( \frac{C_{LR}}{0.1} \right),
\end{align}
where $C_{LR}$ is inside the parenthesis of Eq.~\eqref{eq-delamu}.

For the $\bsll$ anomaly, the Wilson coefficients are given by~\cite{Buras:1994dj,Bobeth:1999mk} 
\begin{align}
 C_9 \sim&\  - \frac{\sqrt{2}}{4G_F}\frac{4\pi}{\alpha_e} \frac{1}{V_{tb}V^*_{ts}}
                 \frac{1}{4v_\Phi^2} (s_R^2+ s_L^2 ) \eps_{Q_2} \eps_{Q_3}, \\
 C_{10} \sim&\  - \frac{\sqrt{2}}{4G_F}\frac{4\pi}{\alpha_e} \frac{1}{V_{tb}V^*_{ts}}
                 \frac{1}{4v_\Phi^2} (s_R^2- s_L^2 ) \eps_{Q_2} \eps_{Q_3},
\end{align}
where the $\Zp$ boson couplings to the SM doublet quarks are parametrized as
\begin{align}
 \left[g^{\Zp}_{d_L}\right]_{ij} \sim  \left[g^{\Zp}_{u_L}\right]_{ij} \sim
       -\gp \eps_{Q_i} \eps_{Q_j}.
\end{align}
Here, $\eps_{Q_i}$ ($i=1,2,3$) is the similar quantity as $s_{L,R}$
and originates from the mixing between the SM and VL quarks,
but we now consider the couplings with the second and third generation quarks
and these are typically small in contrast to that for muons.
The value of $C_9$ is estimated as 
\begin{align}
\label{eq-Cnine}
 C_9 \sim -0.62 \times \left(\frac{1.0~\TeV}{v_\Phi}\right)^2
                       \left(\frac{s_L^2+s_R^2}{1}\right)
                        \left(\frac{\eps_{Q_2}\eps_{Q_3}}{-0.002}\right).
\end{align}

\subsection{$\chi^2$ analysis}

\begin{table}[t]
 \centering
\caption{\label{tab-bestinfo}
Values of $\chi^2$, selected input parameters 
and observables at the best fit points A, B, C and  D.
The degree of freedom in our analysis
is $N_\mathrm{obs}-N_\mathrm{inp} = 98-65 = 33$.
The last column shows the experimental central values and their uncertainties.  
The upper limits on the lepton flavor violating decays are $90\%$ C.L. limits.
}
\small 
 \begin{tabular}{c|cccc|c} \hline 
 Parameters                         & Point A            & Point B    & Point C & Point D      \\   \hline\hline
$\chi^2$                            & $22.6$               & $25.0$    &  $23.3$ & $23.8$       \\ 
$g'$                                   & $0.250$             & $0.340$  & $0.323$& $0.349$     \\
$m_\Zp$ [GeV]                                                          &  277.6 & 535.3 & 486.7 & 758.7     \\ \hline
 Observables                                   & Point A            & Point B  &Point C & Point D          & Exp. \\   \hline\hline
 $\Delta a_\mu \times 10^9$                  & $2.62$ & $2.52$ & $2.52$    & $2.45$           & $2.68\pm0.76$ \\
 $\br{\mu}{e\gamma}\times 10^{13}$  & $0.147$& $1.597$& $0.061$ & $0.822$          &$<4.2$ \\
 $\br{\tau}{\mu\gamma}\times10^{8}$
                                        &$\val{3.34}{-4}$&$\val{3.62}{-4}$&$\val{3.27}{-6}$&$\val{8.45}{-7}$&$<4.4$  \\
 $\br{\tau}{\mu\mu\mu}\times10^{8}$
                                        &$\val{6.96}{-3}$&$\val{4.77}{-4}$&$\val{6.55}{-5}$ &$\val{4.36}{-7}$ &$<2.1$ \\ \hline
 $\mathrm{Re}\,C_9^\mu$               & $-0.548$               & $-0.806$& $-0.838$& $-0.808$ & $-0.7\pm0.3$  \\
 $\mathrm{Re}\,C_{10}^\mu$          & $0.370$                & $0.252$&$0.347$     & $0.322$  & $0.4\pm0.2$  \\
 $\Delta M_d\,[\mathrm{ps}^{-1}]$   & $0.561$               & $0.610$ & $0.598$    & $0.590$ & $0.506\pm 0.081$   \\
 $\Delta M_s\,[\mathrm{ps}^{-1}]$   & $19.6$                & $19.8$    & $19.4$      & $20.0  $  & $17.76\pm 2.5$  \\
\hline
 \end{tabular}
\end{table}

We searched for parameter points which can explain the anomalies 
without unacceptably changing the other observables consistent with the SM. 
The $\chi^2$ function is defined as 
\begin{align}
 \chi^2(x):= \sum_{I \in \mathrm{obs}} \frac{\left( y_I(x) - y_I^0\right)^2}{\sigma_I^2},  
\end{align}
where $x$ is a parameter space point, $y_I(x)$ is 
the value of an observable $I$ 
whose central value is $y_I^0$ and uncertainty is $\sigma_I$.
In this model, there are 65 input parameters 
and we studied 98 observables including fermion masses, CKM matrix, 
lepton and quark flavor violation observables, and so on. 
The full list of observables and the values for the $\chi^2$ fitting 
are shown in Ref.~\cite{Kawamura:2019hxp}. 
Table~\ref{tab-bestinfo} shows values of $\chi^2$, selected input parameters 
and observables at the best fit points A, B, C and  D.
At these points, both anomalies are explained successfully, 
while the flavor violation observables, e.g. $\mu \to e \gamma$ and $\Delta M_s$ 
are consistent with the current limits. 
We did a global analysis for the model, 
and we found that $\damu$ can be explained 
only if $m_\Zp < 800~\GeV$ and $m_{E} < 1.5~\TeV$. 
Therefore, these new particles will be probed by the LHC experiment as discussed 
in the next section. 
Although there are no lower bounds on the branching fractions 
in the lepton flavor violating decays,     
we found the relations among the decay modes 
$\br{\mu}{e\gamma} \gg \br{\mu}{eee}$  
and 
$\br{\tau}{\mu\gamma} \sim \br{\tau}{\mu\mu\mu} \gg \br{\tau}{e\gamma,e\ell\ell}$   
($\ell = e,\mu$) 
which would be confirmed by the future experiments.  
In the quark sector, most observables are consistent with the SM predictions 
assuming unitarity of the CKM matrix, 
although our model can have non-unitarity of the CKM matrix for the SM families.

\section{LHC signals}
\label{sec-LHC}

The VL leptons and $\Zp$ boson are within the reach of the LHC experiment. 
As shown in Eq.~\eqref{eq-Cnine}, 
the $\Zp$ boson typically couples to SM quarks only with small couplings. 
Hence, the strong constraints from the di-muon resonance search~\cite{Aad:2019fac} 
for a $\Zp$ boson can be evaded 
since the production cross section is suppressed 
even though the branching fraction to muons is sizable. 
Therefore, the $\Zp$ boson below $1~\TeV$ is, in general, not excluded by 
the $\Zp$ search.

\begin{figure}[t]
\begin{minipage}[c]{0.32\hsize}
 \centering
    \includegraphics[height=36mm]{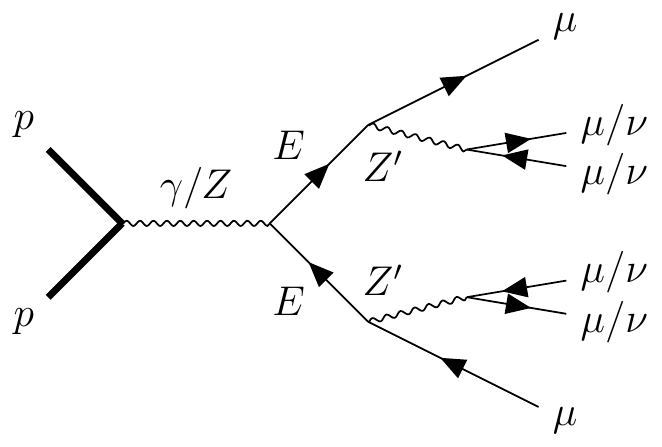}
\end{minipage}
\begin{minipage}[c]{0.32\hsize}
 \centering
    \includegraphics[height=36mm]{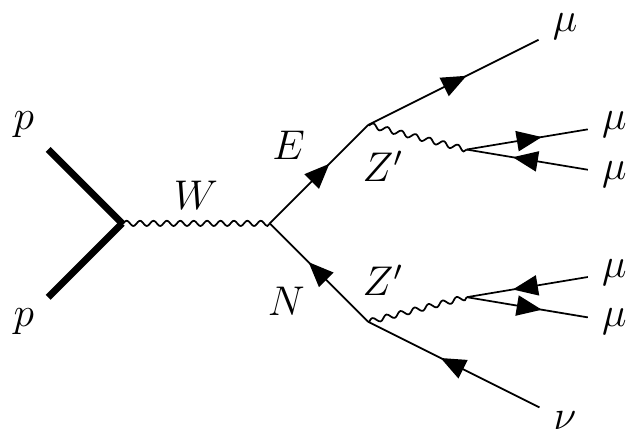}
\end{minipage}
\begin{minipage}[c]{0.32\hsize}
 \centering
    \includegraphics[height=36mm]{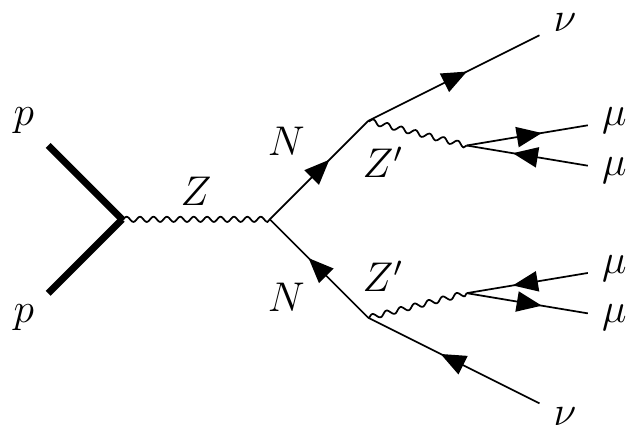}
\end{minipage}
 \caption{\label{fig-proc}
Processes with four muons or more. 
}
\end{figure}

The processes which can produce more than four muons are shown in Fig~\ref{fig-proc}. 
The signal of more than four muons is produced 
if either of the $\Zp$ boson decays to a pair of muons from  
charged VL lepton pair production (left), 
while both have to decay to muons in the other processes (middle and right) 
involving the VL neutrino.

\begin{figure}[t]
 \centering
    \includegraphics[height=0.35\hsize]{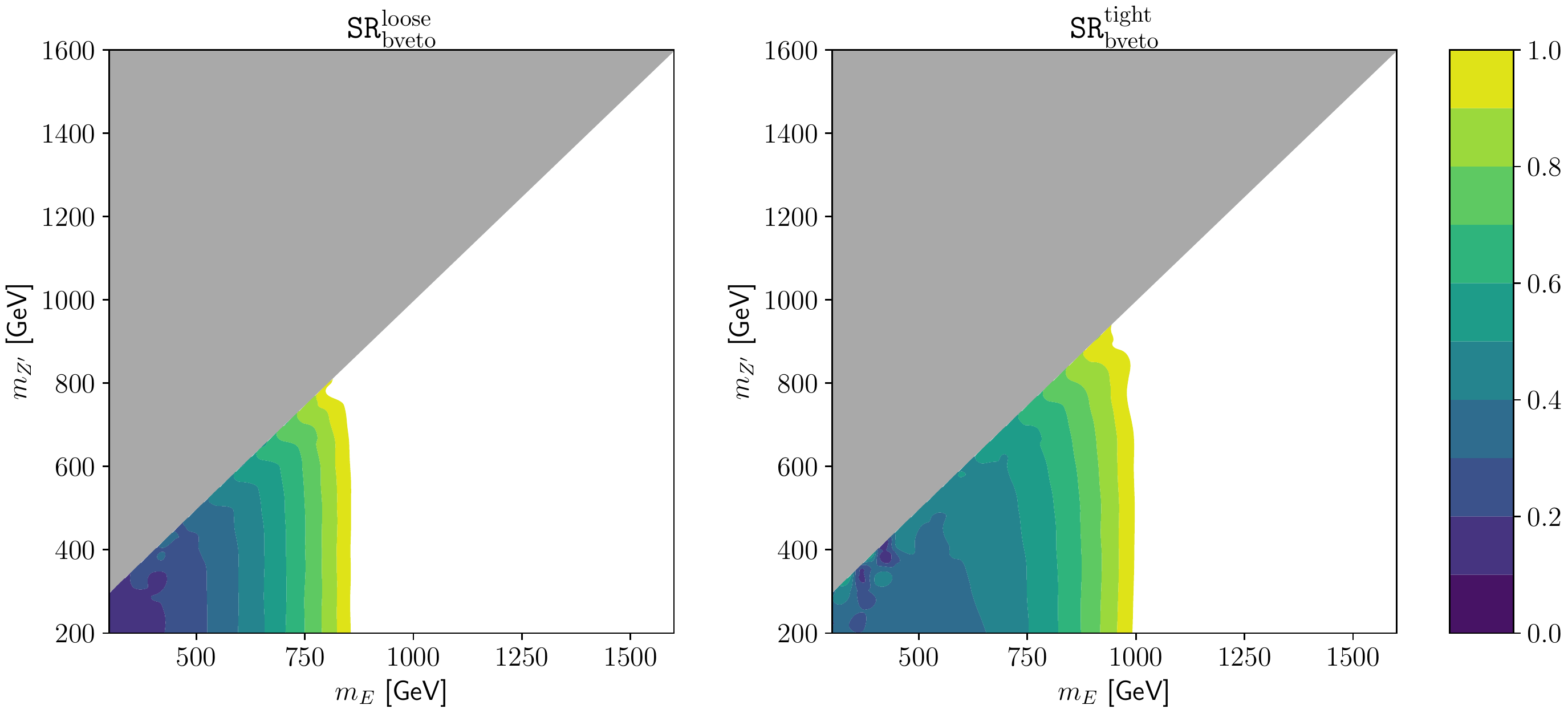}
 \caption{\label{fig-limS}
Limits from the signal regions with 4 leptons for the singlet-like VL lepton. 
The colors signify upper bounds on the $\br{E}{\Zp \mu}$, 
e.g. $\br{E}{\Zp \mu} > 0.5$ is excluded at $(m_E, m_{\Zp}) = (750,200)~\GeV$ 
in the right panel. 
}
\end{figure}

We recast the limits obtained in Refs.~\cite{Aad:2021qct} 
searching for signals with more than four leptons. 
We have generated events using \texttt{MadGraph5$\_$2$\_$8$\_$2}~\cite{Alwall:2014hca}
based on a \texttt{UFO}~\cite{Degrande:2011ua} model file generated
with \texttt{FeynRules$\_$2$\_$3$\_$43}~\cite{Christensen:2008py,Alwall:2014hca}.
The events are showered with \texttt{PYTHIA8}~\cite{deFavereau:2013fsa}
and then run through the fast detector simulator \texttt{Delphes3.4.2}~\cite{deFavereau:2013fsa}.
We used the default ATLAS card for the detector simulation,
but the threshold on $p_T$ for the muon efficiency formula
is changed to $5$ GeV from $10$ GeV
since muons with $p_T>5~\GeV$ are counted as signal muons in Ref.~\cite{Aad:2021qct}.

The current limits for the singlet-like VL lepton are shown in Fig.~\ref{fig-limS}. 
We see that the $\SRtt$ gives the strongest bounds on the $\br{E_1}{\Zp \mu}$. 
Typically, $\br{E_1}{\Zp\mu}\sim 1$ when $m_\chi > m_{E_1} > m_\Zp$, 
while $\br{E_1}{\Zp\mu}\sim \br{E_1}{\chi\mu}\sim 0.5$ 
when $m_{E_1} > m_\Zp, m_\chi$.  
The limit is about $1~\TeV$ $(750~\GeV)$ 
nearly independent of the $\Zp$ mass when the branching fraction is 1 (0.5). 
When the branching fraction is 1, 
the limit for the doublet-like lepton is about 1.3 TeV.

\section{Summary} 
\label{sec-sum}

In this work, we constructed a model with a complete VL fourth family 
and a $\Up$ gauge symmetry. 
The anomaly in the muon anomalous magnetic moment is explained 
by the 1-loop diagrams mediated by the $\Zp$ boson and the VL leptons, 
and those in the rare $B$ decays are explained by the tree-level $\Zp$ boson exchange.  
We searched for good-fit parameter points by a global $\chi^2$ analysis, 
and we found plenty of points that can explain the anomalies. 
At these points, the other observables, 
such as lepton flavor violating decays and neutral meson mixing,  
are consistent with the current limits. 
We then proposed a novel possibility to detect signals with four muons or more at the LHC. 
By recasting the latest data, the current limit 
for the singlet-like (doublet-like) VL lepton is 1.0 (1.3) TeV 
when the $\br{E}{\Zp\mu} = 1$.

\section*{Acknowledgments}
I would like to thank Stuart Raby and Andreas Trautner for collaborating on projects 
related to this talk. 
The work of J.K is supported in part by the Institute for Basic Science (IBS-R018-D1).

{\footnotesize 
\bibliographystyle{JHEP}
\bibliography{reference_vectorlike}

}

\end{document}